# The Fermi Paradox is Neither Fermi's Nor a Paradox

## Robert H. Gray

**Abstract**

The so-called Fermi paradox claims that if technological life existed anywhere else, we would see evidence of its visits to Earth—and since we do not, such life does not exist, or some special explanation is needed. Enrico Fermi, however, never published anything on this topic. On the one occasion he is known to have mentioned it, he asked "where is everybody?"—apparently suggesting that we don't see extraterrestrials on Earth because interstellar travel may not be feasible, but *not* suggesting that intelligent extraterrestrial life does not exist, or suggesting its absence is paradoxical.

The claim "they are not here; therefore they do not exist" was first published by Michael Hart, claiming that interstellar travel and colonization of the galaxy would be inevitable if intelligent extraterrestrial life existed, and taking its absence here as proof that it does not exist anywhere. The Fermi paradox appears to originate in Hart's argument, not Fermi's question.

Clarifying the origin of these ideas is important, because the Fermi paradox is seen by some as an authoritative objection to searching for evidence of extraterrestrial intelligence—cited in the U. S. Congress as a reason for killing NASA's SETI program on one occasion—but evidence indicates that it misrepresents Fermi's views, misappropriates his authority, deprives the actual authors of credit, and is not a valid paradox.

## 1. Introduction

The so-called Fermi paradox is often cited in discussions about the search for extraterrestrial intelligence (SETI), and has generated a sizable literature (Brin, 1983; Webb, 2011). One typical definition is (Horowitz, 2002):

> If technologically advanced civilizations have inhabited our Galaxy for timescales of approximately a billion years, and if some of these have engaged in interstellar travel and colonization, then why have we not seen physical evidence of their visits?

This implies that Enrico Fermi, a great physicist, was skeptical about the possible existence of technologically advanced extraterrestrial life and viewed the absence of visits as a paradox, but the available evidence shows that this was not his view.

Fermi apparently never published a word on the subject of extraterrestrial life or interstellar travel. He is known to have asked "where is everybody?" at a lunch in 1950, questioning the plausibility of interstellar travel according to eye-witness accounts presented later. If Fermi was skeptical about interstellar travel, then using his name for an argument that requires it is misleading.

This paper will show that the phrase "Fermi paradox" first appeared in 1977, 27 years after Fermi's question in 1950, and in connection with Michael Hart's 1975 argument "they are not here; therefore they do not exist," which was extended by Frank Tipler in 1980. The Fermi paradox appears to confuse Fermi's question with the Hart-Tipler argument—misrepresenting Fermi's views and obscuring the identity of the actual authors.

The Fermi paradox caused a crisis in SETI beginning around 1975 (Dick, 1998):

> This crisis undermined the very foundation of the SETI endeavor, bringing into question the logic of the radio search paradigm by claiming that all searches of the electromagnetic spectrum might well be fruitless.

Senator William Proxmire (D-WI) cited Tipler's name when he killed NASA's SETI program in 1981 for the following year (Proxmire, 1981) using language very much like the Fermi paradox:

... if intelligent beings did exist elsewhere and possessed the technology for interstellar communication they would have developed interstellar travel and thus would already be present in our solar system.

NASA's SETI program was restarted but canceled a second time in 1993 (Garber, 2014) by Senator Richard Bryan (D-NV), and since then no U. S. government funds have been appropriated for searches.

This purported paradox attributed to Fermi continues to have influence today, generating academic discussion and possibly contributing to a *de facto* prohibition on government support for research in a branch of astrobiology, but it is clearly not Fermi's and does not constitute a valid paradox.

## 2. Origin of the phrase *Fermi Paradox*

The phrase 'Fermi paradox' seems to have first appeared in print in March of 1977, referring to Fermi's question "where is everybody?" (Stephenson, 1977):

The first and simplest answer to 'Fermi's paradox' has been put forward by Hart ... mankind is the first intelligent species within this Galaxy, and that consequently there could have been no visitations to this Solar System by other civilizations.

Stephenson believes that he coined the term for publication (Stephenson, 2014):

At the time the word paradox was used in conversation to describe the problems facing SETI researchers and Carl Sagan's story about Fermi and "Where is everybody" was often mentioned. I put single quotes around 'Fermi's Paradox' ... as an apology for using a short convenient two word label for a large and indistinct field of conjecture.

The Fermi paradox appeared as the subject of a conference session in April, 1977 (Martin, *et al.*, 1979), and the phrase began appearing frequently in the early 1980s—cited in 19 places in one SETI symposium (Papagiannis, 1984)—and it appears frequently today, five places in one recent book (Vakoch, 2014). The phrase was not found in any publication earlier than Stephenson's paper, and was not found in any title in a SETI bibliography with 1,488 references through February 1977 (Mallove, 1978).

It is surprising that the phrase did not appear in print until 1977—27 years after Fermi's question in 1950, but only two years after Hart's 1975 paper—and in connection with Hart's name, suggesting that the Fermi paradox is more closely related to Hart's argument than Fermi's question.

## 3. Origin of Fermi's Question

Fermi is known to have asked "where is everybody?" during a lunch at Los Alamos in 1950 (Jones, 1985; Finney & Jones, 1985). Fermi's question is often taken as challenging the idea that technological extraterrestrial life exists because we see no evidence of visits, but accounts from people who were present make it clear that Fermi was questioning the feasibility of interstellar travel and *not* questioning the possible existence of technological extraterrestrial life.

Eric Jones solicited letters in 1984 from the three surviving people present at the lunch—Emil Konopinski, Edward Teller, and Herbert York (Fermi died in 1954)—asking them about the occasion. Excerpts from the responses show that the conversation focused on interstellar travel and its feasibility:

Konopinski:

When I joined the party I found being discussed evidence about flying saucers. That immediately brought to my mind a cartoon I had recently seen in the New Yorker, explaining why public trash cans were disappearing from the streets of New York City. ... The cartoon showed what was evidently a flying saucer sitting in the background and, streaming toward it, "little green men" (endowed with antennas) carrying trash cans. ... There ensued a discussion as to whether the saucers could somehow exceed the speed of light.

Teller:

I do not believe that much came from this conversation, except perhaps a statement that the distances to the next location of living beings may be very great and that, indeed, as far as our galaxy is concerned, we are living somewhere in the sticks, far removed from the metropolitan area of the galactic center.

York (about Fermi):

... he went on to conclude that the reason that we hadn't been visited might be that interstellar flight is impossible, or, if it is possible, always judged to be not worth the effort, or technological civilization doesn't last long enough for it to happen.

York clearly recalled Fermi as questioning the feasibility of interstellar travel, Teller seems to have thought that distance (and by implication, difficulty of interstellar travel) was the reason for not seeing extraterrestrials, and Konopinski did not address the topic, although he places the event in time, because such a cartoon appeared May 20, 1950. *None* of the respondents reported that Fermi questioned the possible existence of extraterrestrials, or suggested that not seeing any was a paradox—yet those ideas are the core of the Fermi paradox.

Fermi's question was recalled by Konopinski as "But where is everybody?", by Teller as "Where is everybody?", and by York as "Don't you wonder where everybody is?", which suggests "where is everybody?" as Fermi's likely wording, although respondents may have been influenced by Jones' use of that phrase. The wording has been given by others who were not present as "where are they?" (Sagan, 1963; Sagan & Shklovski, 1966; Oliver & Billingham, 1971; Drake & Sobel, 1992), and some give the time as during WWII, but the eyewitness accounts seem most authoritative.

Fermi's question seems to have first appeared in print in a footnote to a discussion of the possibility that contact with an extraterrestrial civilization occurred within historical times (Sagan, 1963):

> This possibility has been seriously raised before; for example, by Enrico Fermi, on a now rather well-known dinner table discussion at Los Alamos during the Second World War, where he introduced the problem with the words "Where are they?"

The context suggests that Sagan thought Fermi's point was that if extraterrestrials had visited the Earth recently, we should see evidence of it, and since we don't, they have not visited recently. Another early mention of Fermi's question appeared in the *Project Cyclops* report (Oliver & Billingham, 1971), apparently challenging the feasibility of interstellar travel:

> If, on the other hand, interstellar travel is much easier than we predict, we would argue that to maintain radio silence is no real protection, for in this case a galactic survey would not need to depend on beacons. The question to be answered in this case is Enrico Fermi's Where are they?

The so-called Fermi paradox clearly misrepresents Fermi's views, by using his name for an argument that assumes interstellar travel (which he *was* questioning), and because it challenges the possible existence of technological extraterrestrial life (which he was *not* questioning). It might be more accurate to describe "where is everybody?" as *Fermi's question about the feasibility of interstellar travel*.

## 4. The Hart-Tipler Argument

Michael Hart appears to have been the first to publish the argument "they are not here; therefore they do not exist" (Hart, 1975):

> If ... there were intelligent beings elsewhere in our Galaxy, then they would eventually have achieved space travel, and would have explored and colonized the Galaxy, as we have explored and colonized the Earth. However... they are not here; therefore they do not exist.

Hart argued that all other possible explanations for the absence of extraterrestrials on Earth are false, grouping them into four categories: (1) "physical explanations," for example, interstellar travel is infeasible, (2) "sociological explanations," for example, extraterrestrials have chosen not to visit Earth, (3) "temporal explanations," for example, they have not had time to reach the Earth, and (4) the Earth has been visited, but we do not observe them here at present. He rejected all four possible explanations and concluded that only his explanation is correct—they do not exist anywhere in the Galaxy, a matter of ~$10^{11}$ stars—and offered a corollary policy conclusion "an extensive search for radio messages from other civilizations is probably a waste of time and money."

David Viewing published a somewhat similar argument assuming interstellar travel and colonization (Viewing, 1975), but unlike Hart, he concluded that extraterrestrials might exist and that we should search for interstellar probes as well as search for signals.

Frank Tipler extended Hart's ideas, suggesting smart machines rather than biological colonists (Tipler, 1980):

> ... a self-replicating universal constructor with intelligence comparable to the human level ... such a machine combined with present-day rocket technology would make it possible to explore and/or colonize the Galaxy in less than 300 million years...

This was a major extension of Hart's ideas, because Hart mentioned "robots" in only one sentence, as potentially manning spaceships on long voyages carrying frozen zygotes. Tipler's mechanism would presumably operate for hundreds of millions of years, follow orders without question or deviation or evolution, explore or colonize billions of planets and perhaps even moons and asteroids at no extra cost, operate in places where life could not survive, and solve a multitude of other problems.

Tipler's conclusion was even more sweeping than Hart's—that Man is probably the only intelligent species in the Universe (Tipler, 1981), not just our Galaxy—a matter of ~$10^{11}$ more stars. That universal result seems strikingly similar to the result of the Genesis story which describes the creation of only one intelligent species (Leeming, 2010), and it is surprising that Tipler did not comment on this similarity, because elsewhere he explained biblical stories such as miracles using physics (Tipler, 2007).

Hart and Tipler's arguments have been challenged on grounds such as the energy required for interstellar travel (Oliver, 1994) and risks of self-replicating machines and rates of colonization (Sagan & Newman 1983). Others have taken their arguments as serious objections to the idea that technological extraterrestrial life might exist elsewhere (Proxmire, 1981, Zuckerman & Hart 1982, Papagiannis, 1985).

The Hart-Tipler argument is very similar to the Fermi paradox, although there is a slight difference. The Fermi paradox is often posed as a question about why we don't see extraterrestrials, using Hart's argument that we should, but as a question it permits answers other than "they do not exist"—although it begs that answer.

## 5. Proper Attribution

Using Fermi's name for the so-called Fermi paradox is clearly mistaken, because (1) it misrepresents Fermi's views, which were skeptical about interstellar travel, but not about the possible existence of extraterrestrials, and (2) its central idea "they are not here; therefore they do not exist" was first published by Hart. Priority of publication and accuracy suggests using a name like *Hart-Tipler argument* instead of "Fermi paradox."

A name change has been suggested before. Charles Seeger suggested "... 'Fermi Paradox' is an unfortunate and thoroughly misleading appellation." (Seeger, 1985). Iosif Shklovsky reportedly urged "... we must call it the 'Hart Paradox,' rather than the 'Fermi Paradox' ..." (Papagiannis, 1985). The phrase "Fermi-Hart paradox" has been used in some papers (Wesson, 1990). Stephen Webb suggested "Tsiolkovsky-Fermi-Viewing-Hart paradox" (Webb, 2002).

Konstantin Tsiolkovsky is said to have recognized similar issues in the 1930s. He believed that space travel was possible, and also believed that life must exist on planets around many other stars, and he noticed potential contradictions (Lytkin, *et al.*, 1995):

(1) if these beings exist they would have visited earth
(2) if they exist they would have given us some sign of their existence

He resolved the contradiction with the idea that "it is not yet time" for them to visit us, and that "our means are too weak to perceive these signs." Lytkin *et al.* suggested that the "Fermi Paradox" would more properly be known as the "Fermi Question" and did not suggest adding Tsiolkovsky's name.

## 6. No Paradox

The word paradox is usually taken to mean a self-contradictory statement, or a seemingly logical argument that leads to a contradiction, and the contradiction suggests that something is wrong.

The Hart-Tipler argument takes the seemingly obvious fact *they are not here* as evidence that a premise "technological extraterrestrials exist" must be false, because if they did exist, their colonization argument leads to the conclusion *they are here*, which seems absurd. This is a *reductio ad absurdum* argument, not a paradox, although like a paradox it depends on every statement being true—yet the argument consists of many speculations which are not known to be true. For example, "colonizing the galaxy" tacitly assumes: (1) interstellar travel is feasible, (2) the Galaxy would be filled quickly, (3) the Earth would be among $\sim 10^{11}$ or more colonized worlds, (4) this entirely colonized state would persist for many millions or billions of years. Any or all of those might be false, so a Hart-Tipler argument or a Fermi paradox making those assumptions can not be valid. Two hundred years ago, *they are not here* could have been said about dinosaurs, although a great deal of evidence has been found since then. The claim of paradox has been rejected before (Freitas, 1983).

A variation on the Hart-Tipler argument is *they do not exist because we have not received their signals* (Wesson, 1990; Wikipedia, 2014). This tacitly assumes that a complete search for signals has been performed, which is not true. The literature on searches (Tarter, 1995) indicates that only a small fraction of the radio spectrum has been searched—0.3 GHz in surveys covering much of the sky (Leigh & Horowitz, 2000) using transit observations, and 2 GHz in targeted searches of 800 stars (Backus, *et al.*, 2004)—out of a terrestrial microwave window from 1-10 GHz, a free-space window up to 60 GHz (Oliver & Billingham, 1971), and much more electromagnetic spectrum beyond including optical. Few searches would have detected low-duty-cycle signals anticipated by some (Benford, 2008; Gray, 2011), because both radio and optical surveys typically observe positions for only minutes. An incomplete search for signals can not be used as evidence of complete absence of technological extraterrestrials.

## 7. Conclusions

The so-called Fermi paradox misrepresents Fermi's views about the feasibility of interstellar travel and the possible existence of intelligent extraterrestrial life, uses his name and authority for ideas originated by Hart and Tipler, and asserts a logical paradox where none exists, so it is difficult to see any valid use for the phrase. It's not Fermi's idea, and it's not a paradox.

Fermi asked "where is everybody?", questioning the feasibility of interstellar travel, but not questioning the possible existence of intelligent extraterrestrial life, so it seems clear that Fermi's name should not be used for a so-called Fermi paradox which depends on interstellar travel, and which does question the existence of extraterrestrials—the opposite of his views. Fermi's question might more accurately be called *Fermi's question about the feasibility of interstellar travel,* to avoid mistaking it as an argument against the possible existence of intelligent extraterrestrial life.

The argument "they are not here; therefore they do not exist*"* was first published by Hart and extended by Tipler, and might be called the *Hart-Tipler argument against the existence of technological extraterrestrials.* This is not exactly the same as the Fermi paradox, but it is the simplest answer to the question "where is everybody?" if interstellar travel and colonization are assumed. It seems misleading to cloak the Hart-Tipler argument with Fermi's name and authority, because doing so deprives the true authors of credit and attributes views to Fermi which he did not hold.

Some people may feel that the so-called Fermi paradox is a sleeping dog that should be left to lie, because it is established in the scientific literature and public mind, but most people would agree that clearly mistaken and misleading terminology should be corrected. The issue is important, because the Hart-Tipler argument was cited as a reason for killing NASA's SETI program on one occasion in the U. S. Congress, and under the guise of Fermi's name and the claim of a logical paradox, it may continue to inhibit funding and research in that area of astrobiology.

**Acknowledgements**

The author is grateful to Amir R. Alexander, Steven J. Dick, Steven D. Lord, and Michael A. Warner for comments on drafts of this paper, and grateful to Eric M. Jones, David G. Stephenson, and Stephen Webb for helpful information.

**Author Disclosure Statement**

No competing financial interests exist.

Address correspondence to:
*Robert H. Gray*
*Gray Consulting*
*Chicago, Illinois*
*E-mail: RobertHansenGray@gmail.com*